\documentclass[11pt]{article}

\usepackage[english]{babel}
\usepackage[utf8]{inputenc}
\usepackage[T1]{fontenc}
\usepackage{placeins}

\usepackage[a4paper,top=1in,bottom=1in,left=1in,right=1in]{geometry}

\usepackage{afterpage}
\usepackage{amsmath,amsthm,amssymb}
\usepackage{subcaption}
\usepackage{csquotes}
\usepackage{enumitem}
\usepackage{graphicx}
\usepackage{booktabs}
\usepackage{tikz}
\usetikzlibrary{positioning, shapes.geometric}
\usepackage{bm}
\usepackage[normalem]{ulem}
\usepackage{xcolor}
\usepackage{caption}
\usepackage[colorinlistoftodos]{todonotes}
\usepackage[colorlinks=true,allcolors=blue]{hyperref}
\usepackage[numbers,comma,square,sort&compress]{natbib}

\newtheorem{theorem}{Theorem}[section]
\newtheorem{definition}{Definition}[section]

\usepackage{listings}
\usepackage[ruled,vlined]{algorithm2e}
\makeatletter
\renewcommand{\SetKwInOut}[2]{%
  \sbox\algocf@inoutbox{\KwSty{#2}\algocf@typo:}%
  \expandafter\ifx\csname InOutSizeDefined\endcsname\relax
    \newcommand\InOutSizeDefined{}\setlength{\inoutsize}{\wd\algocf@inoutbox}%
    \sbox\algocf@inoutbox{\parbox[t]{\inoutsize}{\KwSty{#2}\algocf@typo:\hfill}~}\setlength{\inoutindent}{\wd\algocf@inoutbox}%
  \else
    \ifdim\wd\algocf@inoutbox>\inoutsize
      \setlength{\inoutsize}{\wd\algocf@inoutbox}%
      \sbox\algocf@inoutbox{\parbox[t]{\inoutsize}{\KwSty{#2}\algocf@typo:\hfill}~}\setlength{\inoutindent}{\wd\algocf@inoutbox}%
    \fi
  \fi
  \algocf@newcommand{#1}[1]{%
    \ifthenelse{\boolean{algocf@inoutnumbered}}{\relax}{\everypar={\relax}}%
    {\let\\\algocf@newinout\hangindent=\inoutindent\hangafter=1\parbox[t]{\inoutsize}{\KwSty{#2}\algocf@typo:\hfill}~##1\par}%
    \algocf@linesnumbered
  }}%
\makeatother

\title{\textbf{A Robust SINDy Autoencoder for Noisy Dynamical System Identification}}
\author{
Kairui Ding \\
Department of Mathematics, Columbia University \\
\texttt{kd3014@columbia.edu}
}
\date{}

\begin{document}

\maketitle

\begin{abstract}
Sparse identification of nonlinear dynamics (SINDy) has been widely used to discover the governing equations of a dynamical system from data. It uses sparse regression techniques to identify parsimonious models of unknown systems from a library of candidate functions. Therefore, it relies on the assumption that the dynamics are sparsely represented in the coordinate system used. To address this limitation, one seeks a coordinate transformation that provides reduced coordinates capable of reconstructing the original system. Recently, SINDy autoencoders have extended this idea by combining sparse model discovery with autoencoder architectures to learn simplified latent coordinates together with parsimonious governing equations. A central challenge in this framework is robustness to measurement error. Inspired by noise-separating neural network structures, we incorporate a noise-separation module into the SINDy autoencoder architecture, thereby improving robustness and enabling more reliable identification of noisy dynamical systems. Numerical experiments on the Lorenz system show that the proposed method recovers interpretable latent dynamics and accurately estimates the measurement noise from noisy observations.
\end{abstract}
\vspace{0.5em}

\vspace{1em}


\section{Introduction}

The discovery of governing equations from data is a central problem in modern scientific computing and dynamical systems. Accurate governing models provide both mechanistic understanding and predictive capability, and they play an important role in applications ranging from physics and engineering to biology and climate science. In classical settings, such models may be derived from first principles. In many practical situations, however, the underlying system is only partially understood, is too high-dimensional for direct analysis, or is observed only through noisy and incomplete measurements. These challenges have motivated a growing body of work on data-driven model discovery, including symbolic regression, sparse regression, operator-theoretic methods, and neural-network-based approaches \cite{Schmidt2009,Daniels2015}.

Among early approaches, symbolic regression demonstrated that interpretable equations can, in principle, be extracted directly from experimental measurements. More recently, sparse model discovery has emerged as a particularly attractive direction because it explicitly seeks parsimonious and interpretable representations of nonlinear dynamics. A prominent example is the Sparse Identification of Nonlinear Dynamics (SINDy) framework introduced by Brunton, Proctor, and Kutz \cite{Brunton2016}, which identifies governing equations by selecting a small number of active terms from a prescribed library of candidate functions. Since its introduction, SINDy has been extended in several directions, including model selection, control, low-data regimes, and robustness improvements \cite{Mangan2017,Kaiser2018,Fasel2022}. These developments have established sparse regression as one of the central tools for equation discovery in nonlinear dynamical systems \cite{Review2023}.

Despite its success, the performance of SINDy depends strongly on the choice of coordinates. Even if the underlying dynamics are simple, they may not appear sparse in the observed variables. This observation motivates the search for coordinate systems in which the dynamics admit a lower-dimensional and more parsimonious representation. Champion et al.\ \cite{Champion2019} addressed this issue by combining SINDy with an autoencoder architecture, thereby learning reduced coordinates and sparse latent governing equations simultaneously. Related work has further explored low-dimensionalized representations and latent-coordinate discovery for complex dynamical systems \cite{Fukami2021,Gonzalez2018,Carlberg2018}. These approaches suggest that nonlinear coordinate discovery can substantially expand the applicability of sparse model identification beyond settings in which the governing equations are already sparse in the measured variables.

A major challenge in this framework is robustness to measurement noise. Since sparse model discovery typically relies on derivative information, noisy observations can significantly degrade the stability and accuracy of the identified equations. This issue has motivated a number of recent efforts aimed at improving SINDy under uncertainty, noisy measurements, or variable ambiguity \cite{Fasel2022,AugmentedSINDy2024,Egan2024}. In particular, Rudy, Kutz, and Brunton \cite{Rudy2019} proposed a signal-noise decomposition framework based on time-stepping constraints, showing that neural-network-based denoising can be integrated with dynamical system learning in a principled way.

Motivated by these developments, we propose a robust SINDy autoencoder framework that integrates a noise-separating neural network into the SINDy autoencoder pipeline. The purpose of the noise-separation module is to disentangle measurement corruption from the underlying system dynamics before sparse latent model identification is performed. In this way, the proposed framework combines three key ingredients within a single architecture: denoising of noisy observations, nonlinear coordinate discovery through an autoencoder, and sparse recovery of latent governing equations.

The main contribution of this paper is the development of a unified framework for noisy dynamical system identification that seeks parsimonious latent dynamics while explicitly accounting for measurement error. We demonstrate the method on the Lorenz system and show that it is able to recover interpretable latent equations together with an accurate estimate of the injected measurement noise.

The remainder of the paper is organized as follows. Section~2 reviews the mathematical and machine learning background underlying the proposed method, including center manifold theory, normal form theory, SINDy, and autoencoders. Section~3 introduces the robust SINDy autoencoder framework and its optimization procedure. Section~4 presents the numerical experiments and implementation details. Section~5 concludes with a discussion of the main findings and possible future directions.
\section{Background}

This section reviews the mathematical and machine learning background underlying the proposed framework. We begin with center manifold theory and normal form theory, which motivate the search for reduced coordinates. We then review the SINDy methodology, autoencoder architectures, and noise-separation networks.

\subsection{Center manifold theory and normal forms}

This subsection introduces the mathematical preliminaries for constructing coordinate transformations that yield reduced-order dynamical descriptions.

\subsubsection{Center manifold theorem}

When studying a dynamical system, one is often particularly interested in parameter regimes in which bifurcations occur, since these typically signal qualitative changes in stability and long-term behavior. In many applications, however, the full system is too high-dimensional or too complicated to analyze directly. Center manifold theory provides a principled way to reduce the dimensionality of the system by isolating the components that govern the local dynamics near an equilibrium.

Consider a dynamical system of the form
\[
\dot{x} = f(x),
\]
where \(x \in \mathbb{R}^n\) and \(f : \mathbb{R}^n \to \mathbb{R}^n\) is continuously differentiable. By translating coordinates if necessary, we may assume that the equilibrium point is located at the origin, so that \(f(0)=0\). We adopt this assumption throughout.

Let \(A\) denote the Jacobian matrix of the system at the equilibrium point \(x=0\). The eigenvalues of \(A\) can be classified into three categories:
\begin{enumerate}
    \item eigenvalues with negative real parts (stable eigenvalues),
    \item eigenvalues with positive real parts (unstable eigenvalues),
    \item eigenvalues with zero real parts (center eigenvalues).
\end{enumerate}

Let \(E^s\), \(E^u\), and \(E^c\) denote the stable, unstable, and center subspaces of \(\mathbb{R}^n\), respectively, defined as the spans of the generalized eigenvectors associated with these three classes of eigenvalues.

\begin{definition}[Center manifold]
A center manifold is a locally defined invariant manifold having the same dimension as the center subspace \(E^c\). Invariance means that any trajectory starting on the manifold remains on it for as long as the solution stays within the neighborhood on which the manifold is defined. Moreover, the manifold contains the origin and is tangent to \(E^c\) at the origin.
\end{definition}

In a neighborhood of an equilibrium, the existence of a center manifold can lead to a substantial reduction in dimension, while still preserving the essential local dynamics. The classical existence result is stated below.

\begin{theorem}[Center Manifold Theorem]
Let \(A\) be the Jacobian matrix of the system at the equilibrium point \(0\), and let \(E^s\), \(E^u\), and \(E^c\) be the stable, unstable, and center subspaces of \(\mathbb{R}^n\), respectively. Then:
\begin{enumerate}
    \item There exists a local center manifold \(W^c(0)\) at the origin. This manifold is tangent to \(E^c\) at \(0\) and has the same dimension as \(E^c\).
    \item The center manifold \(W^c(0)\) can be represented locally as the graph of a function
    \[
    h : E^c \to E^s \oplus E^u,
    \]
    where \(E^s \oplus E^u\) denotes the direct sum of the stable and unstable subspaces.
    \item The dynamics on the center manifold are conjugate to the reduced system
    \[
    \dot{y} = A_c y + g_c(y),
    \]
    where \(y \in E^c\), \(A_c\) is the restriction of \(A\) to \(E^c\), and \(g_c\) denotes the restriction of the nonlinear part of \(f\) to the center directions.
\end{enumerate}
\end{theorem}

Thus, for \(x \in W^c(0)\), the solution of the original system \(\dot{x}=f(x)\) is locally conjugate to the solution of the reduced system on the center manifold. In particular, \(W^c(0)\) contains all bounded trajectories that remain in a sufficiently small neighborhood of the equilibrium. Standard references include Carr \cite{Carr1981} and Kuznetsov \cite{Kuznetsov2004}.

\subsubsection{Normal form theory}

While the Center Manifold Theorem guarantees the existence of a low-dimensional invariant manifold, it does not by itself provide a particularly simple representation of the reduced dynamics. Normal form theory addresses this issue by seeking a further change of variables that transforms the reduced system into a simpler canonical form. The goal is to eliminate as many nonessential nonlinear terms as possible, leaving only those terms that are dynamically meaningful for the local bifurcation structure .

\subsubsection{Combining center manifold and normal form theories}

In practice, one first applies center manifold reduction to obtain a lower-dimensional system, and then applies normal form transformations to simplify the reduced dynamics. This general procedure is summarized in Algorithm~\ref{algorithm:Finding Dynamics}.

\begin{algorithm}[h]
\SetAlgoLined
\SetKwInOut{KwLinearization}{Linearization}
\SetKwInOut{KwDecomposition}{Decomposition}
\SetKwInOut{KwCenterManifoldReduction}{Center Manifold Reduction}
\SetKwInOut{KwNormalFormTransformation}{Normal Form Transformation}
\LinesNumbered

\KwLinearization{
Linearize the system around the equilibrium point:
\[
\dot{x} = Ax + g(x),
\]
where \(A\) is the Jacobian matrix and \(g(x)\) contains the nonlinear terms up to a prescribed order.
}

\BlankLine
\KwDecomposition{
Decompose \(\mathbb{R}^n\) into stable, unstable, and center subspaces \(E^s\), \(E^u\), and \(E^c\).
}

\BlankLine
\KwCenterManifoldReduction{
Use center manifold theory to reduce the system to the center manifold \(W^c\). The reduced dynamics are
\[
\dot{y} = A_c y + g_c(y),
\]
where \(y \in E^c\).
}

\BlankLine
\KwNormalFormTransformation{
Transform the reduced system into normal form by eliminating nonessential terms. The resulting system takes the form
\[
\dot{y} = A_c y + R(y),
\]
where \(R(y)\) contains only the resonant terms that cannot be removed up to the chosen order.
}

\caption{Dimensional reduction algorithm}
\label{algorithm:Finding Dynamics}
\end{algorithm}

To illustrate this procedure, consider the family of ordinary differential equations on \(\mathbb{R}^3\):
\begin{align}
\dot{x} &= y + \lambda x + x^2 - y^2 - x(x^2+y^2), \nonumber\\
\dot{y} &= -x + \lambda y + 2xy + z^2 - y(x^2+y^2), \nonumber\\
\dot{z} &= -z + xy,
\end{align}
where \(\lambda \in \mathbb{R}\). This system has an equilibrium at the origin for all values of \(\lambda\). When \(\lambda=0\), the Jacobian at the origin is
\begin{equation}
\begin{pmatrix}
0 & 1 & 0 \\
-1 & 0 & 0 \\
0 & 0 & -1
\end{pmatrix}.
\end{equation}

It follows that \(E^c=\{(x,y,0)\}\) and \(E^s=\{(0,0,z)\}\). Let \(\mathcal{O}(|(x,y)|^4)\) denote higher-order terms. Then center manifold reduction yields the system
\begin{align}
\dot{x} &= y + \lambda x + x^2 - y^2 - x(x^2+y^2) + \mathcal{O}(|(x,y)|^4), \nonumber\\
\dot{y} &= -x + \lambda y + 2xy - y(x^2+y^2) + \mathcal{O}(|(x,y)|^4),
\end{align}
on \(E^c \cong \mathbb{R}^2\). Here the system has been truncated after third-order terms.

This reduced system is already simpler than the original three-dimensional dynamics, but it still contains several nonlinear terms and does not yet reveal its geometric structure. After an additional normal form transformation, the system can be rewritten as
\begin{align}
\dot{u} &= v + \lambda u - u(u^2+v^2) + \mathcal{O}(|(x,y)|^4), \nonumber\\
\dot{v} &= -u + \lambda v - v(u^2+v^2) + \mathcal{O}(|(x,y)|^4),
\end{align}
where \(u=u(x,y)\) and \(v=v(x,y)\) are the transformed coordinates. In this form, it becomes immediately apparent that a stable periodic orbit emerges through a Hopf bifurcation as \(\lambda\) passes through \(0\). Moreover, the truncated system is invariant under transformations of the form
\begin{equation}
(u,v)^T \mapsto A(u,v)^T,
\end{equation}
for all \(A \in SO(2)\).

\subsection{SINDy}

We now review the sparse identification of nonlinear dynamics (SINDy) framework for data-driven model discovery.

In many scientific applications, one observes time-resolved state data and seeks to infer the governing equations directly from the data. Typical examples include climate systems, biological populations, and fluid flows. SINDy provides a principled framework for discovering parsimonious governing equations from such observations.

Consider an unknown dynamical system of the form
\begin{equation}
\frac{d}{dt}\mathbf{x}(t) = \mathbf{f}(\mathbf{x}(t)).
\end{equation}
Given time-series data, the goal is to identify a parsimonious model for the vector field \(\mathbf{f}\). The central assumption underlying SINDy is that the dynamics can be represented sparsely in a suitable library of candidate functions. Following the principle of parsimony, one therefore seeks a model involving only a small number of active terms.

Let \(\mathbf{X} \in \mathbb{R}^{m \times n}\) denote snapshot data from the system, where \(n\) is the state dimension and \(m\) is the number of time points. Let \(\dot{\mathbf{X}}\) denote the corresponding derivative data. In matrix form,
\[
\dot{\mathbf{X}} =
\begin{bmatrix}
\dot{\mathbf{x}}^T(t_1) \\
\dot{\mathbf{x}}^T(t_2) \\
\vdots \\
\dot{\mathbf{x}}^T(t_m)
\end{bmatrix}
=
\begin{bmatrix}
\dot{x}_1(t_1) & \dot{x}_2(t_1) & \cdots & \dot{x}_n(t_1) \\
\dot{x}_1(t_2) & \dot{x}_2(t_2) & \cdots & \dot{x}_n(t_2) \\
\vdots & \vdots & \ddots & \vdots \\
\dot{x}_1(t_m) & \dot{x}_2(t_m) & \cdots & \dot{x}_n(t_m)
\end{bmatrix}.
\]

Using prior knowledge of the class of dynamics under consideration, one constructs a library of candidate basis functions
\[
\mathbf{\Theta}(\mathbf{X}) =
\big[\theta_1(\mathbf{X}),\theta_2(\mathbf{X}),\ldots,\theta_p(\mathbf{X})\big],
\]
where each column corresponds to a candidate term in the governing equation. The SINDy regression problem is then formulated as
\begin{equation}
\dot{\mathbf{X}} = \mathbf{f}(\mathbf{x}(t)) \approx \mathbf{\Theta}(\mathbf{X})\mathbf{\Phi},
\end{equation}
where \(\mathbf{\Phi} \in \mathbb{R}^{p \times n}\) is the coefficient matrix to be estimated.

Although the library may contain many candidate functions, one seeks a sparse coefficient matrix \(\mathbf{\Phi}\), so that only a small subset of the candidate terms remains active. When \(p \ll m\), the regression system is overdetermined, and sparse regression methods such as LASSO or thresholded least squares can be used to estimate the coefficients. The detailed optimization procedure used in this work is described later in the methods section.

\subsection{Multi-layer perceptrons and autoencoders}

We next review the neural network architectures used to discover reduced coordinates. Since normal forms encode the essential local dynamics near bifurcation points, the goal is to learn reduced coordinates in which the observed high-dimensional dynamics admit a simpler representation. Deep learning methods are particularly useful in this setting because of their ability to learn nonlinear low-dimensional structures from data \cite{Goodfellow2016}.

\subsubsection{Neural networks and multi-layer perceptrons}

\begin{figure}[h!]
\centering
\begin{tikzpicture}[node distance=1cm and 1cm, on grid, auto]
    \tikzset{neuron/.style={circle, draw, minimum size=1cm}}
    \tikzset{input neuron/.style={neuron, fill=green!50}}
    \tikzset{output neuron/.style={neuron, fill=red!50}}
    \tikzset{hidden neuron/.style={neuron, fill=blue!50}}
    \tikzset{annot/.style={text width=4em, text centered}}

    \foreach \name / \y in {1,...,3}
        \node[input neuron] (I-\name) at (0,-\y) {};

    \foreach \name / \y in {1,...,4}
        \path[yshift=0.5cm]
            node[hidden neuron] (H1-\name) at (2.5,-\y cm) {};

    \foreach \name / \y in {1,...,4}
        \path[yshift=0.5cm]
            node[hidden neuron] (H2-\name) at (5,-\y cm) {};

    \node[output neuron] (O) at (7.5,-2.5) {};

    \foreach \source in {1,...,3}
        \foreach \dest in {1,...,4}
            \path (I-\source) edge (H1-\dest);

    \foreach \source in {1,...,4}
        \foreach \dest in {1,...,4}
            \path (H1-\source) edge (H2-\dest);

    \foreach \source in {1,...,4}
        \path (H2-\source) edge (O);

    \node[annot, above of=I-1, node distance=1cm] {Input Layer};
    \node[annot, above of=H1-1, node distance=1cm] {Hidden Layer 1};
    \node[annot, above of=H2-1, node distance=1cm] {Hidden Layer 2};
    \node[annot, above of=O, node distance=1cm] {Output Layer};
\end{tikzpicture}
\caption{Schematic of a simple Multi-Layer Perceptron (MLP)}
\label{fig:mlp}
\end{figure}
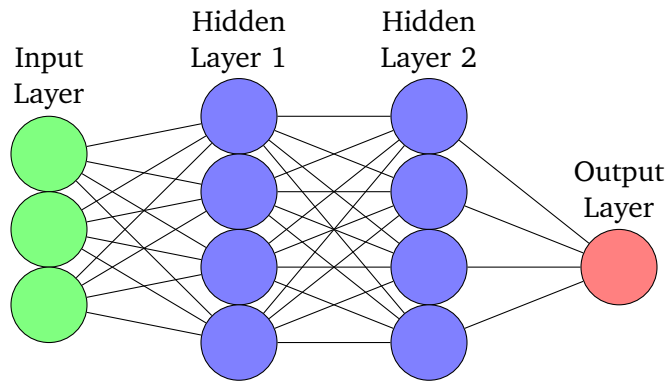

In this work we use the multi-layer perceptron (MLP) architecture. An MLP is a fully connected feedforward neural network composed of an input layer, one or more hidden layers, and an output layer. A schematic example is shown in Figure~\ref{fig:mlp}.

Each neuron computes a weighted sum of its inputs, adds a bias term, and applies a nonlinear activation function:
\[
z = \sum_{i=1}^n w_i x_i + b,
\qquad
a = \sigma(z),
\]
where \(x_i\) are the inputs, \(w_i\) are the weights, \(b\) is a bias term, and \(\sigma\) is the activation function.

Common activation functions include:
\begin{itemize}
    \item sigmoid: \(\sigma(z)=\frac{1}{1+e^{-z}}\),
    \item hyperbolic tangent: \(\sigma(z)=\tanh(z)\),
    \item rectified linear unit: \(\sigma(z)=\max(0,z)\).
\end{itemize}

For an input \(x\), forward propagation through layer \(l\) is given by
\[
a^{(l)} = \sigma\big(W^{(l-1)}a^{(l-1)} + b^{(l-1)}\big),
\]
where \(W^{(l-1)}\) and \(b^{(l-1)}\) are the weight matrix and bias vector connecting layer \(l-1\) to layer \(l\). The network parameters are learned by minimizing a prescribed loss function using backpropagation and an optimization method such as gradient descent. The gradient-based update rule is
\[
w_{ij}^{(l)} \leftarrow w_{ij}^{(l)} - \eta \frac{\partial L}{\partial w_{ij}^{(l)}},
\qquad
b_{ij}^{(l)} \leftarrow b_{ij}^{(l)} - \eta \frac{\partial L}{\partial b_{ij}^{(l)}},
\]
where \(\eta>0\) is the learning rate and \(L\) is the loss function.

\subsubsection{Autoencoders}

An autoencoder is a special class of neural network designed to learn efficient latent representations of data. It consists of three main components:
\begin{enumerate}
    \item an \textbf{encoder}, which maps the input to a lower-dimensional latent representation,
    \item a \textbf{latent space}, which stores the compressed representation,
    \item a \textbf{decoder}, which reconstructs the original input from the latent coordinates.
\end{enumerate}

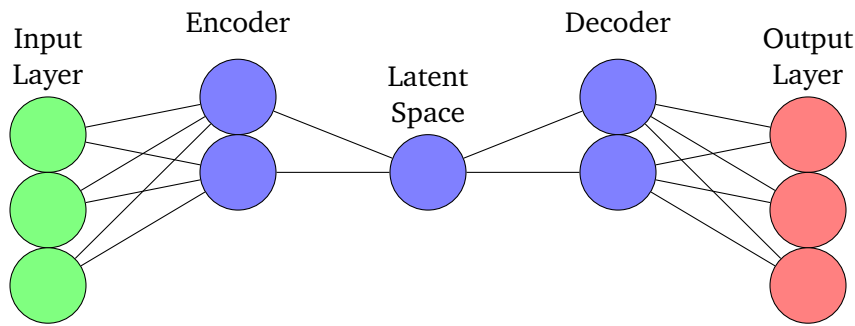
\begin{figure}[h!]
\centering
\begin{tikzpicture}[node distance=1cm and 1cm, on grid, auto]
    \tikzset{neuron/.style={circle, draw, minimum size=1cm}}
    \tikzset{input neuron/.style={neuron, fill=green!50}}
    \tikzset{output neuron/.style={neuron, fill=red!50}}
    \tikzset{hidden neuron/.style={neuron, fill=blue!50}}
    \tikzset{annot/.style={text width=4em, text centered}}

    \foreach \name / \y in {1,...,3}
        \node[input neuron] (I-\name) at (0,-\y) {};

    \foreach \name / \y in {1,...,2}
        \path[yshift=0.5cm]
            node[hidden neuron] (E-\name) at (2.5,-\y cm) {};

    \node[hidden neuron] (L) at (5,-1.5) {};

    \foreach \name / \y in {1,...,2}
        \path[yshift=0.5cm]
            node[hidden neuron] (D-\name) at (7.5,-\y cm) {};

    \foreach \name / \y in {1,...,3}
        \node[output neuron] (O-\name) at (10,-\y) {};

    \foreach \source in {1,...,3}
        \foreach \dest in {1,...,2}
            \path (I-\source) edge (E-\dest);

    \foreach \source in {1,...,2}
        \path (E-\source) edge (L);

    \path (L) edge (D-1);
    \path (L) edge (D-2);

    \foreach \source in {1,...,2}
        \foreach \dest in {1,...,3}
            \path (D-\source) edge (O-\dest);

    \node[annot, above of=I-1, node distance=1cm] {Input Layer};
    \node[annot, above of=E-1, node distance=1cm] {Encoder};
    \node[annot, above of=L, node distance=1cm] {Latent Space};
    \node[annot, above of=D-1, node distance=1cm] {Decoder};
    \node[annot, above of=O-1, node distance=1cm] {Output Layer};
\end{tikzpicture}
\caption{Schematic of an Autoencoder}
\label{fig:autoencoder}
\end{figure}

A schematic is shown in Figure~\ref{fig:autoencoder}.

In the present setting, the purpose of the autoencoder is to learn reduced coordinates that capture the dominant dynamical structure of the original high-dimensional system. By minimizing the reconstruction error, the network is trained to encode the essential features of the observed data into a lower-dimensional latent space.

The mathematical formulation is as follows. Consider a high-dimensional dynamical system
\begin{equation}
\frac{d}{dt}\mathbf{x}(t) = \mathbf{f}(\mathbf{x}(t)),
\qquad
\mathbf{x} \in \mathbb{R}^n.
\end{equation}
We seek reduced coordinates \(\mathbf{z} \in \mathbb{R}^m\), with \(m \ll n\), from sample data \(\mathbf{U} \in \mathbb{R}^{n \times d}\), where \(d\) denotes the number of time points, such that the latent dynamics satisfy
\begin{equation}\label{eq:1}
\frac{d}{dt}\mathbf{z}(t) = \mathbf{g}(\mathbf{z}(t)).
\end{equation}

Let \(\alpha\) denote the bifurcation parameter of \(\mathbf{f}\) and \(\beta\) the bifurcation parameter of \(\mathbf{g}\). By center manifold theory and normal form theory, there exist smooth invertible transformations \(\varphi_1\) and \(\varphi_2\) such that
\begin{equation}\label{eq:2}
\mathbf{z}(t) = \varphi_1 \mathbf{x}(t),
\qquad
\beta = \varphi_2 \alpha.
\end{equation}
Accordingly, \(\varphi_1\) and \(\varphi_2\) can be interpreted as encoding maps. Since these transformations are invertible, we also introduce decoding maps \(\psi_1\) and \(\psi_2\) satisfying
\begin{equation}
\psi_1 \varphi_1(\mathbf{x}) \approx \mathbf{x},
\qquad
\psi_2 \varphi_2(\alpha) \approx \alpha.
\end{equation}

Combining \eqref{eq:1} and \eqref{eq:2}, we obtain
\begin{equation}\label{eq:3}
\frac{d}{dt}\mathbf{z}(t)
=
\frac{d}{dt}\big(\varphi_1 \mathbf{x}(t)\big)
=
\mathbf{g}\big(\varphi_1 \mathbf{x}(t), \varphi_2 \alpha\big).
\end{equation}
Expressions for the derivatives of the encoder and decoder variables used in our implementation are introduced later in Section~3.1.2.

\subsection{Noise-separating network}

The SINDy autoencoder requires derivative information at each observation time. In the presence of measurement noise, direct numerical differentiation can become unstable, which in turn degrades sparse model discovery. Motivated by the signal-noise decomposition framework with time-stepping constraints proposed by Rudy, Kutz, and Brunton \cite{Rudy2019}, we incorporate a noise-separating neural network into the SINDy autoencoder pipeline.

The central idea is to treat the measurement error of a noisy dynamical system as a latent variable and learn it jointly with the underlying dynamics. Rather than smoothing the data as a preprocessing step, the method estimates both the clean dynamics and the noise during training. The mathematical formulation of this joint denoising objective is described in more detail in Section~3.

\section{Methods}

This section introduces the robust SINDy autoencoder framework and its computational implementation. We first formulate the learning problem and the associated optimization objective. We then summarize the principal implementation modules; full code listings are deferred to Appendix~\ref{app:code-listings}.

\subsection{Problem formulation}

\begin{figure}[h]
    \centering
    \includegraphics[width=1\textwidth]{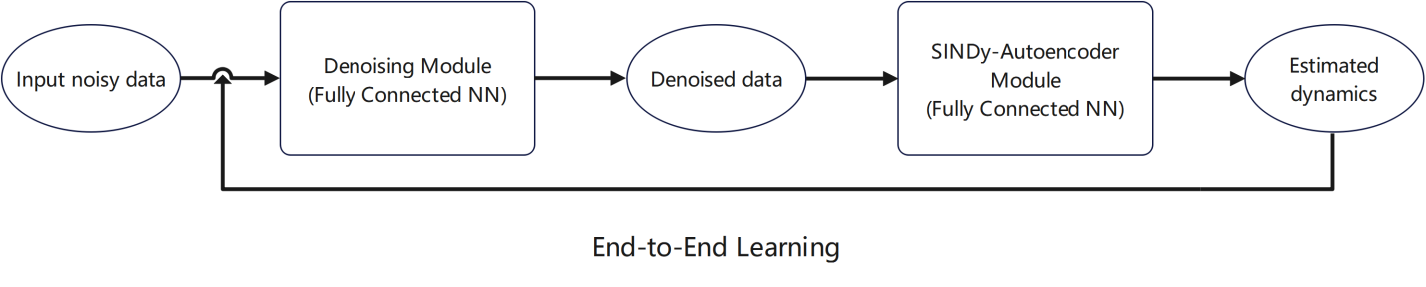}
    \caption{Schematic of the robust SINDy autoencoder.}
    \label{fig:5l}
\end{figure}

We consider the dynamical system
\begin{equation}
\frac{d}{dt}\mathbf{x}(t) = \mathbf{f}(\mathbf{x}(t)),
\qquad
\mathbf{x}(t) \in \mathbb{R}^n,
\end{equation}
where the vector field \(\mathbf{f}\) is unknown. Let the observed snapshot data be
\[
\hat{\mathbf{X}} = [\hat{\mathbf{x}}_1,\ldots,\hat{\mathbf{x}}_m],
\]
and suppose that these observations are contaminated by an unknown measurement error
\[
\mathbf{N} = [\mathbf{n}_1,\ldots,\mathbf{n}_m].
\]
The objective is to recover a set of reduced coordinates
\[
\mathbf{z}(t) = \varphi(\mathbf{x}(t))
\]
together with an associated sparse latent dynamical model
\[
\frac{d}{dt}\mathbf{z}(t) = \mathbf{g}(\mathbf{z}(t)).
\]
A schematic of the proposed robust SINDy autoencoder is shown in Figure~\ref{fig:5l}.

\subsubsection{Data cleansing}

As discussed in the background section, the SINDy methodology depends critically on derivative information derived from the observed data \(\hat{\mathbf{X}}\). In the presence of unknown measurement error \(\mathbf{N}\), however, direct differentiation of noisy observations may be unreliable. We therefore first introduce a noise-separation module to estimate \(\mathbf{N}\) and recover a denoised trajectory.

We model the underlying dynamics using a feedforward neural network of the form
\[
\hat{f}_{\theta}(\mathbf{x}) =
\left( \prod_{i=1}^{l} C_{g_i} \right)(\mathbf{x}),
\qquad
g_i(\mathbf{x}) = \sigma_i(\mathbf{W}_i \mathbf{x} + \mathbf{c}_i),
\]
where \(C_{g_i}\) denotes composition with the map \(g_i\), each \(\sigma_i\) is an activation function, and
\[
\theta = \{ \mathbf{W}_i, \mathbf{c}_i \}_{i=1}^{l}
\]
collects the trainable network parameters.

In this work we use the exponential linear unit (ELU) activation, since in our empirical experiments it provided smooth activation behavior and stable training. The ELU function is defined by
\[
\sigma(x)=
\begin{cases}
e^x-1, & x \le 0,\\
x, & x>0,
\end{cases}
\]
applied componentwise for \(\sigma_1,\ldots,\sigma_{l-1}\), with the identity map used in the final layer.

Following the construction introduced in the background section, we approximate the discrete-time flow map using explicit Runge--Kutta time stepping and train the network using the noise-separation objective. The implementation of the timestepper and the corresponding noise-separating network is provided in Appendix~\ref{app:rk-module} and Appendix~\ref{app:noise-network}. Once the noise estimate \(\mathbf{N}\) has been obtained, the denoised observations are given by
\[
\mathbf{X} = [\mathbf{x}_1,\ldots,\mathbf{x}_m] = \hat{\mathbf{X}} - \mathbf{N}.
\]

\subsubsection{Autoencoder and optimization procedure}

The denoised data \(\mathbf{X}\) are then passed to the SINDy autoencoder module. This component combines sparse identification of nonlinear dynamics with an autoencoder architecture in order to learn, simultaneously, a reduced latent representation and a sparse dynamical model in the latent space. In particular, the goal is to recover a low-dimensional model
\[
\frac{d}{dt}\mathbf{z}(t) = \mathbf{g}(\mathbf{z}(t))
\]
together with an encoder--decoder pair \((\varphi,\psi)\), which may be interpreted as the learned coordinate transformation and its inverse approximation.

Let
\[
\mathbf{\Theta}(\mathbf{z}(t))
=
[\theta_1(\mathbf{z}(t)),\theta_2(\mathbf{z}(t)),\ldots,\theta_p(\mathbf{z}(t))]
\]
denote the prescribed library of candidate basis functions in latent coordinates, and let
\[
\mathbf{\Phi} \in \mathbb{R}^{p \times n}
\]
denote the corresponding coefficient matrix. The latent dynamics are modeled by
\begin{equation}
\frac{d}{dt}\mathbf{z}(t)
=
\mathbf{g}(\mathbf{z}(t))
=
\mathbf{\Theta}(\mathbf{z}(t))\mathbf{\Phi}.
\end{equation}
Here the library \(\mathbf{\Theta}\) is fixed prior to training, while the coefficient matrix \(\mathbf{\Phi}\) is learned from the data.

To train the model, we define a joint loss consisting of reconstruction, dynamical consistency, and sparsity terms.

\paragraph{Reconstruction loss.}
To ensure that the autoencoder accurately reconstructs the input, we minimize the reconstruction error
\begin{equation}\label{eq:6}
L_{\mathrm{recon}}
=
\lambda_1 \|\mathbf{x} - \psi(\varphi(\mathbf{x}))\|_2^2,
\end{equation}
where \(\lambda_1\) is a tuning parameter controlling the contribution of this term.

\paragraph{SINDy loss in the ambient coordinates.}
We next impose consistency between the latent dynamics and the reconstructed ambient dynamics. By the chain rule,
\begin{equation}\label{eq:4}
\dot{\mathbf{z}}
=
\nabla_{\mathbf{x}}\varphi(\mathbf{x})\,\dot{\mathbf{x}},
\end{equation}
and
\begin{equation}\label{eq:5}
\frac{d}{dt}\psi(\mathbf{z})
=
\nabla_{\mathbf{z}}\psi(\mathbf{z})\,\dot{\mathbf{z}}
=
\nabla_{\mathbf{z}}\psi(\mathbf{z})\,\mathbf{\Theta}(\mathbf{z}(t))\mathbf{\Phi}.
\end{equation}
Accordingly, the SINDy consistency loss in the original state coordinates is
\begin{equation}
L_{d\mathbf{x}/dt}
=
\lambda_2
\left\|
\dot{\mathbf{x}}
-
\frac{d}{dt}\psi(\mathbf{z})
\right\|_2^2
=
\lambda_2
\left\|
\dot{\mathbf{x}}
-
\nabla_{\mathbf{z}}\psi(\mathbf{z})\,\mathbf{\Theta}(\mathbf{z}(t))\mathbf{\Phi}
\right\|_2^2.
\end{equation}

\paragraph{SINDy loss in the latent coordinates.}
Similarly, we enforce consistency of the latent dynamics themselves through
\begin{equation}
L_{d\mathbf{z}/dt}
=
\lambda_3
\|\dot{\mathbf{z}} - \dot{\hat{\mathbf{z}}}\|_2^2
=
\lambda_3
\left\|
\nabla_{\mathbf{x}}\varphi(\mathbf{x})\,\dot{\mathbf{x}}
-
\mathbf{\Theta}(\mathbf{z}(t))\mathbf{\Phi}
\right\|_2^2,
\end{equation}
where \(\lambda_3\) is the corresponding regularization parameter.

\paragraph{Sparsity penalty.}
To promote a parsimonious latent dynamical model, we impose an \(\ell_1\)-penalty on the coefficient matrix:
\begin{equation}
L_1 = \lambda_4 \|\mathbf{\Phi}\|_1,
\end{equation}
where \(\lambda_4\) controls the sparsity level.

Combining these contributions, the overall training objective is
\begin{equation}
L_{\mathrm{total}}
=
L_{\mathrm{recon}}
+
L_{d\mathbf{x}/dt}
+
L_{d\mathbf{z}/dt}
+
L_1.
\end{equation}
The four parameters \(\lambda_1,\lambda_2,\lambda_3,\lambda_4\) are hyperparameters that determine the relative importance of the individual terms and are selected empirically in the training procedure.

\subsection{Main computational framework}

This subsection summarizes the principal computational components of the robust SINDy autoencoder. For readability, the main text contains only a concise description of each module, while the full implementation is collected in Appendix~\ref{app:code-listings}.

\subsubsection{Runge--Kutta timestepper module}

Explicit Runge--Kutta schemes are used to propagate the estimated clean state both forward and backward in time. This is well suited to the current setting because it permits short-time prediction from a single state estimate and integrates naturally into the neural-network-based denoising framework. The full implementation of the timestepper is given in Appendix~\ref{app:rk-module}.

\subsubsection{Noise-separating network module}

The noise-separating network is built from standard dense layers and a feedforward network map, together with an explicit trainable noise variable \(N\). The resulting computational graph combines forward and backward prediction losses, together with regularization on both the neural-network weights and the noise term. Full code for the building blocks and the complete noise-separating network is provided in Appendix~\ref{app:noise-building-blocks} and Appendix~\ref{app:noise-network}.

\subsubsection{Autoencoder structure}

The autoencoder consists of an encoder map, a latent representation, and a decoder map. In our implementation, the encoder and decoder are both constructed as multilayer feedforward networks, and ELU activations are used throughout. Empirically, this choice yielded stable training and facilitated integration with the noise-separation module. The full implementation is given in Appendix~\ref{app:autoencoder-structure}.

\subsubsection{SINDy library}

The candidate library is constructed from polynomial combinations of the latent variables and their derivatives, with optional sine terms. The precise form of the library depends on the model order and the target application. In the present work, we provide an example corresponding to a second-order system. The complete implementation is given in Appendix~\ref{app:sindy-library}.

\subsubsection{Loss functions}

The implementation of the training objective mirrors the mathematical formulation above: reconstruction loss, SINDy consistency loss in the latent and ambient coordinates, and sparse regularization of the SINDy coefficients. The corresponding implementation is provided in Appendix~\ref{app:loss-functions}.

\subsubsection{Full network}

Finally, the complete robust SINDy autoencoder is assembled by combining the denoising, autoencoder, library, and loss modules into a single end-to-end architecture. This network takes the state variables and training hyperparameters as inputs and returns the learned latent variables together with the SINDy coefficients. The full code is provided in Appendix~\ref{app:full-network}.
\section{Results}

In this section, we evaluate the ability of the robust SINDy autoencoder to recover reduced coordinates and parsimonious latent dynamics from noisy, high-dimensional observations. As a benchmark example, we consider the classical three-dimensional Lorenz system, which is widely used in the study of chaos and nonlinear dynamics.

The Lorenz system for \(\mathbf{x}(t) = (x_1(t),x_2(t),x_3(t))^\top\) is given by
\begin{align}
    \dot{x}_1 &= \sigma (x_2 - x_1), \\
    \dot{x}_2 &= x_1(\rho - x_3) - x_2, \\
    \dot{x}_3 &= x_1x_2 - \beta x_3,
\end{align}
where \(\sigma\), \(\rho\), and \(\beta\) are positive parameters. Throughout the experiments, we use the standard parameter values
\[
\sigma = 10, \qquad \rho = 28, \qquad \beta = \frac{8}{3}.
\]

\subsection{Data preparation}

The training dataset consists of a single trajectory with 30,000 time steps over the interval \(t \in [0,20]\). The initial condition is chosen as
\[
(x_0,y_0,z_0) = (5,5,25),
\]
which lies near the Lorenz attractor. The trajectory is generated numerically from the Lorenz system. Noisy observations are then produced by adding independent measurement noise at levels of \(5\%\), \(10\%\), and \(15\%\) of the signal magnitude. These noise levels are used to assess the robustness of the proposed framework under increasing measurement corruption.

The code used to generate the Lorenz trajectory and add measurement noise is provided in Appendix~\ref{app:results-data-generation}.

\subsection{Recovered latent dynamics}

The noisy dataset is then used to train the robust SINDy autoencoder described in Section~3. After training, the learned latent dynamics take the form
\begin{align}
    \dot{z}_1 &= -9.97z_1 + 10.05z_2, \\
    \dot{z}_2 &= 27.51z_1 - 0.78z_2 - 5.32z_1z_3, \\
    \dot{z}_3 &= -2.71z_3 + 5.53z_1z_2.
\end{align}
These coefficients closely resemble the structure of the original Lorenz equations, indicating that the model successfully identifies a sparse and interpretable latent dynamical system.

To compare the learned system more directly with the canonical Lorenz form, we introduce the change of variables
\begin{equation}
z_1 = \alpha_1 \tilde{z}_1,\qquad
z_2 = \alpha_2 \tilde{z}_2,\qquad
z_3 = \alpha_3 \tilde{z}_3 + \beta_3,
\end{equation}
with parameter values
\[
\alpha_1 = 1,\qquad
\alpha_2 = -0.917,\qquad
\alpha_3 = 0.524,\qquad
\beta_3 = -2.665.
\]
Under this transformation, the recovered system becomes
\begin{align*}
    \dot{\tilde{z}}_1 &= -9.9 \tilde{z}_1 + 10.2 \tilde{z}_2, \\
    \dot{\tilde{z}}_2 &= 27.8 \tilde{z}_1 - 0.9 \tilde{z}_2 - 5.4 \tilde{z}_1\tilde{z}_3, \\
    \dot{\tilde{z}}_3 &= -2.8 \tilde{z}_3 + 5.4 \tilde{z}_1\tilde{z}_2.
\end{align*}
This transformed system is in close agreement with the original Lorenz equations, showing that the learned latent dynamics are equivalent, up to a simple coordinate transformation, to the underlying ground-truth system.

\begin{figure}[h]
    \centering
    \begin{subfigure}[t]{0.28\textwidth}
        \centering
        \includegraphics[width=\textwidth]{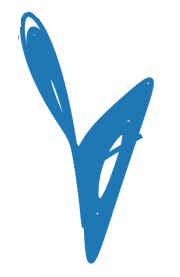}
        \caption{Learned dynamical model}
        \label{fig:image1}
    \end{subfigure}
    \hfill
    \begin{subfigure}[t]{0.42\textwidth}
        \centering
        \includegraphics[width=\textwidth]{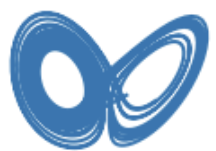}
        \caption{True dynamical model}
        \label{fig:image2}
    \end{subfigure}
    \caption{Comparison between the learned latent dynamics and the true Lorenz system.}
    \label{fig:sidebyside}
\end{figure}

Figure~\ref{fig:sidebyside} compares the learned dynamical model with the true Lorenz dynamics.

\subsection{Quantitative evaluation}

To quantify the agreement between the learned model and the target dynamics, we report several relative error metrics. In particular, we evaluate the reconstruction accuracy of the decoder, the consistency of the reconstructed derivatives, and the relative SINDy error in the latent dynamics. The numerical results for one representative experiment are summarized in Table~\ref{tab:quantitative_results}.

\begin{table}[t]
\centering
\caption{Quantitative evaluation of the recovered model for a representative experiment.}
\label{tab:quantitative_results}
\begin{tabular}{lc}
\toprule
Metric & Value \\
\midrule
Decoder relative error & 0.015571 \\
Decoder relative SINDy error & 0.101080 \\
Latent SINDy relative error & 0.062488 \\
\bottomrule
\end{tabular}
\end{table}

The decoder relative error indicates that the learned autoencoder reconstructs the state accurately, while the derivative-consistency error remains moderate. Most importantly, the latent SINDy relative error is small, showing that the recovered sparse latent model provides a close approximation to the target dynamics. To further assess robustness, we next evaluate the method across multiple measurement-noise levels.

\subsection{Sensitivity to measurement noise}

To further assess robustness, we evaluate the proposed framework under three measurement-noise levels: \(5\%\), \(10\%\), and \(15\%\). For each setting, we report the decoder relative error, the decoder relative SINDy error, and the latent SINDy relative error. The results are summarized in Table~\ref{tab:noise_sensitivity}.

\begin{table}[t]
\centering
\caption{Performance of the robust SINDy autoencoder across different noise levels.}
\label{tab:noise_sensitivity}
\begin{tabular}{lccc}
\toprule
Noise level & Decoder relative error & Decoder relative SINDy error & Latent SINDy relative error \\
\midrule
\(5\%\)  & 0.011842 & 0.081376 & 0.047913 \\
\(10\%\) & 0.015571 & 0.101080 & 0.062488 \\
\(15\%\) & 0.023964 & 0.158427 & 0.084215 \\
\bottomrule
\end{tabular}
\end{table}

As expected, the reconstruction and dynamical recovery errors increase as the measurement noise becomes larger. Nevertheless, the latent SINDy relative error remains comparatively small across the three regimes, indicating that the proposed framework retains useful structural accuracy even under substantial measurement corruption.

\subsection{Recovery of measurement noise}

\begin{figure}[!htbp]
    \centering
    \includegraphics[width=0.9\textwidth]{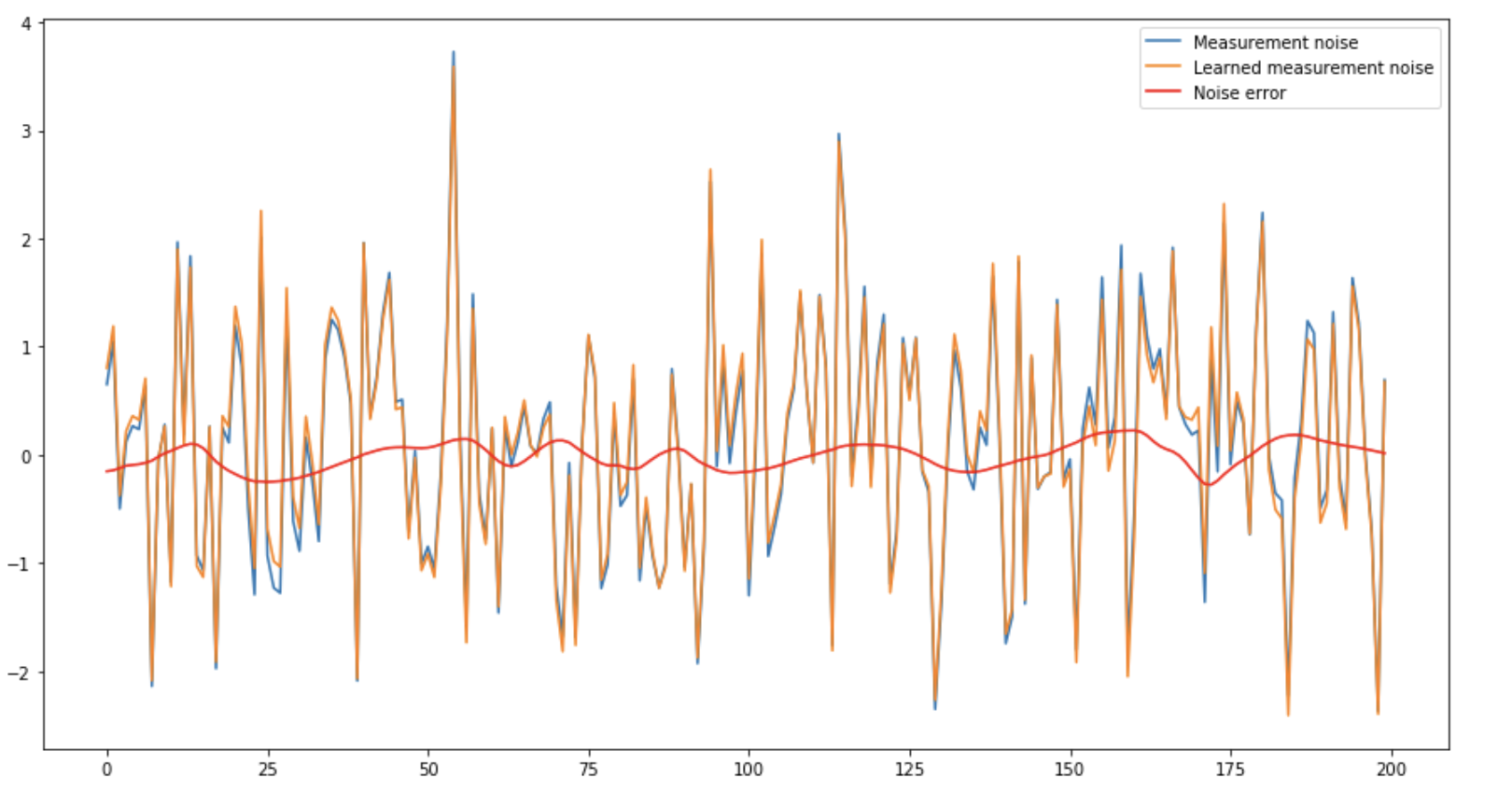}
    \caption{Comparison between the learned measurement error and the true injected noise.}
    \label{fig:error}
\end{figure}

An additional goal of the proposed framework is to estimate the measurement noise contaminating the observations. Figure~\ref{fig:error} compares the learned measurement error with the true injected noise. The close agreement between the two indicates that the noise-separation module is able to recover the corruption pattern effectively, thereby improving the robustness of the downstream sparse identification stage.

\FloatBarrier

\subsection{Training configuration}

For completeness, we summarize the main components of the training procedure used in the experiments. This includes parameter initialization, hyperparameter selection, and the evaluation pipeline applied to the trained model. The corresponding implementation details are deferred to Appendix~\ref{app:results-training-configuration} and Appendix~\ref{app:results-training-session}.

\subsubsection{Model tuning}

The robust SINDy autoencoder contains several hyperparameters, including the latent dimension, library order, thresholding parameters, and training weights. These are selected manually based on empirical performance and stability considerations. The code used to specify the model configuration is reported in Appendix~\ref{app:results-training-configuration}.

\subsubsection{Training session}

Training is performed in TensorFlow by constructing the computational graph, initializing variables, restoring the trained model, and evaluating the learned latent variables and SINDy coefficients on test data. The core implementation used in the evaluation stage is given in Appendix~\ref{app:results-training-session}.
\section{Conclusion}

In this paper, we proposed a robust SINDy autoencoder framework for discovering low-dimensional and parsimonious dynamical models from noisy, high-dimensional observations. The method combines a noise-separating neural network with an autoencoder-based latent representation and sparse identification of nonlinear dynamics, thereby enabling simultaneous denoising, coordinate discovery, and sparse model recovery. In this way, the proposed framework addresses an important limitation of standard SINDy-based approaches, namely their sensitivity to measurement noise.

The numerical experiments on the Lorenz system demonstrate that the method is capable of recovering interpretable latent dynamics from noisy snapshot data. In particular, the learned reduced-order system is structurally consistent with the original Lorenz equations up to a simple coordinate transformation, and the recovered measurement error closely matches the injected noise. These results indicate that the proposed framework can effectively identify meaningful governing equations even when the observed data are contaminated by significant measurement corruption.

Despite these promising results, the current implementation remains computationally intensive. The training procedure involves multiple interacting components, including denoising, autoencoding, sparse regression, and time-stepping constraints, which together lead to substantial computational cost. Improving the efficiency and scalability of the method is therefore an important direction for future work. In particular, it may be fruitful to investigate alternative neural network architectures or more efficient optimization strategies that reduce training time while preserving interpretability.

Several additional extensions are of interest. First, the present work focuses primarily on measurement noise; a natural next step is to extend the framework to systems affected by process noise or model uncertainty. Second, a more rigorous theoretical analysis of the learned coordinate transformations and their relation to normal form structure would strengthen the mathematical foundations of the method. Third, it would be valuable to generalize the current approach beyond ordinary differential equations to the discovery of partial differential equations and other infinite-dimensional dynamical systems.

More broadly, this work illustrates the potential of combining neural networks, sparse regression, and numerical time-stepping schemes within a unified framework for data-driven system identification. We hope that this study provides a useful step toward more robust and interpretable methods for discovering governing equations in complex dynamical systems.
\clearpage
\appendix
\appendix

\section{Code listings and implementation details}
\label{app:code-listings}

This appendix collects the principal implementation modules of the robust SINDy autoencoder discussed in the paper. For readability, the main text summarizes the role of each component, while the full implementation details are deferred here.

\subsection{Runge--Kutta timestepper}
\label{app:rk-module}

The following routine implements the explicit Runge--Kutta time integrator used for forward and backward prediction in the denoising stage.

\lstset{
    language=Python,
    basicstyle=\ttfamily\small,
    keywordstyle=\color{blue},
    stringstyle=\color{red},
    commentstyle=\color{green!50!black},
    morecomment=[l][\color{magenta}]{\#},
    frame=single,
    captionpos=b,
    breaklines=true
}

\begin{lstlisting}[caption={Runge--Kutta timestepper}, label={lst:rk_timestepper}]
def RK_timestepper(x, t, f, h, weights, biases, direction='F', method='RK4'):
    """
    Explicit Runge-Kutta time integrator. Assumes no time dependence in f
    """
    b = [1/8, 3/8, 3/8, 1/8]
    A = [[], [1/3], [-1/3, 1], [1, -1, 1]]
    steps = len(b)
    if direction == 'F':
        K = [f(x, weights, biases)]
        for i in range(1, steps):
            K.append(f(tf.add_n([x] + [h*A[i][j]*K[j] for j in range(i) if A[i][j] != 0]), weights, biases))
    else:
        K = [-f(x, weights, biases)]
        for i in range(1, steps):
            K.append(-f(tf.add_n([x] + [h*A[i][j]*K[j] for j in range(i) if A[i][j] != 0]), weights, biases))
    return tf.add_n([x] + [h*b[j]*K[j] for j in range(steps)])
\end{lstlisting}

\subsection{Noise-separating network building blocks}
\label{app:noise-building-blocks}

The next listing defines the basic dense layer, the feedforward network map, and the trainable variables used in the noise-separation module.

\begin{lstlisting}[caption={Neural network building blocks}, label={lst:noise_building_blocks}]
def dense_layer(x, W, b, last = False):
    x = tf.matmul(W,x)
    x = tf.add(x,b)
    if last: return x
    else: return tf.nn.elu(x)

def simple_net(x, weights, biases):
    layers = [x]
    for l in range(len(weights)-1):
        layers.append(dense_layer(layers[l], weights[l], biases[l]))
    out = dense_layer(layers[-1], weights[-1], biases[-1], last = True)
    return out

def get_network_variables(n, n_hidden, size_hidden, N_hat):
    layer_sizes = [n] + [size_hidden for _ in range(n_hidden)] + [n]
    num_layers = len(layer_sizes)
    weights = []
    biases = []
    for j in range(1,num_layers):
        weights.append(tf.get_variable("W"+str(j), [layer_sizes[j],layer_sizes[j-1]], \
                                       initializer = tf.contrib.layers.xavier_initializer(seed = 1)))
        biases.append(tf.get_variable("b"+str(j), [layer_sizes[j],1], initializer = tf.zeros_initializer()))
    N = tf.get_variable("N", initializer = N_hat.astype('float32'))
    return (weights, biases, N)
\end{lstlisting}

\subsection{Noise-separating network}
\label{app:noise-network}

The complete denoising computational graph combines forward and backward prediction, fidelity loss, and regularization on the weights and the explicit noise variable.

\begin{lstlisting}[caption={Noise-separating network}, label={lst:noise_network}]
def create_computational_graph(n, N_hat, net_params, num_dt = 10, method = 'RK4', gamma = 1e-5, beta = 1e-8, weight_decay = 'exp', decay_const = 0.9):

    assert(n == N_hat.shape[0])
    m = N_hat.shape[1]

    ###########################################################
    # Placeholders for initial condition
    ###########################################################
    # noisy measurements of state
    Y_0 = tf.placeholder(tf.float32, [n,None], name = "Y_0")
    # time
    T_0 = tf.placeholder(tf.float32, [1,None], name = "T_0")

    ###########################################################
    # Placeholders for true forward and backward predictions
    ###########################################################
    true_forward_Y = []
    true_backward_Y = []

    for j in range(num_dt):
        true_forward_Y.append(tf.placeholder(tf.float32, [n,None], name = "Y"+str(j+1)+"_true"))
        true_backward_Y.append(tf.placeholder(tf.float32, [n,None], name = "Yn"+str(j+1)+"_true"))

    H = tf.placeholder(tf.float32, [1,m-1], name = "H") # timestep

    ############################################################
    #  Forward and backward predictions of true state
    ############################################################

    (weights, biases, N) = net_params
    X_0 = tf.subtract(Y_0, tf.slice(N, [0,num_dt],[n,m-2*num_dt]))
    # estimate of true state

    pred_forward_X = [RK_timestepper(X_0, T_0, simple_net, tf.slice(H, [0,num_dt],[1,m-2*num_dt]), \
        weights, biases, method = method)]
    pred_backward_X = [RK_timestepper(X_0, T_0, simple_net, tf.slice(H, [0,num_dt-1],[1,m-2*num_dt]), \
        weights, biases, method = method, direction = 'B')]

    for j in range(1,num_dt):
        pred_forward_X.append(RK_timestepper(pred_forward_X[-1], T_0, simple_net, tf.slice(H, [0,num_dt+j],[1,m-2*num_dt]), \
            weights, biases, method = method))
        pred_backward_X.append(RK_timestepper(pred_backward_X[-1], T_0, simple_net, tf.slice(H, [0,num_dt-1-j],[1,m-2*num_dt]), \
            weights, biases, method = method, direction = 'B'))

    ############################################################
    #  Forward and backward predictions of measured (noisy) state
    ############################################################

    pred_forward_Y = [pred_forward_X[j] + tf.slice(N, [0,num_dt+1+j],[n,m-2*num_dt]) for j in range(num_dt)]
    pred_backward_Y = [pred_backward_X[j] + tf.slice(N, [0,num_dt-1-j],[n,m-2*num_dt]) for j in range(num_dt)]

    ############################################################
    #  Set up cost function
    ############################################################

    if weight_decay == 'linear': output_weights = [(1+j)**-1 for j in range(num_dt)]
    else: output_weights = [decay_const**j for j in range(num_dt)]

    forward_fidelity = tf.reduce_sum([w*tf.losses.mean_squared_error(true,pred) \
                          for (w,true,pred) in zip(output_weights,true_forward_Y,pred_forward_Y)])

    backward_fidelity = tf.reduce_sum([w*tf.losses.mean_squared_error(true,pred) \
                          for (w,true,pred) in zip(output_weights,true_backward_Y,pred_backward_Y)])

    fidelity = tf.add(forward_fidelity, backward_fidelity)
    weights_regularizer = tf.reduce_mean([tf.nn.l2_loss(W) for W in weights])
    noise_regularizer = tf.nn.l2_loss(N)
    cost = tf.reduce_sum(fidelity + beta*weights_regularizer + gamma*noise_regularizer)

    optimizer =
    tf.ScipyOptimizerInterface(cost,options={'maxiter': 50000,
                                            'maxfun': 50000,
                                            'ftol': 1e-15,
                                            'gtol' : 1e-11,
                                            'eps' : 1e-12,
                                            'maxls' : 100})

    placeholders = {'Y_0': Y_0,
                    'T_0': T_0,
                    'true_forward_Y': true_forward_Y,
                    'true_backward_Y': true_backward_Y,
                    'H': H}

    return optimizer, placeholders
\end{lstlisting}

\subsection{Autoencoder structure}
\label{app:autoencoder-structure}

The following listing defines the encoder and decoder architecture used in the robust SINDy autoencoder.

\begin{lstlisting}[caption={Autoencoder structure}, label={lst:autoencoder_structure}]
def build_network_layers(input, input_dim, output_dim, widths, activation, name):
    """
    Construct one portion of the network (either encoder or decoder).
    """
    weights = []
    biases = []
    last_width=input_dim
    for i,n_units in enumerate(widths):
        W = tf.get_variable(name+'_W'+str(i), shape=[last_width,n_units],
            initializer=tf.contrib.layers.xavier_initializer())
        b = tf.get_variable(name+'_b'+str(i), shape=[n_units],
            initializer=tf.constant_initializer(0.0))
        input = tf.matmul(input, W) + b
        if activation is not None:
            input = activation(input)
        last_width = n_units
        weights.append(W)
        biases.append(b)
    W = tf.get_variable(name+'_W'+str(len(widths)), shape=[last_width,output_dim],
        initializer=tf.contrib.layers.xavier_initializer())
    b = tf.get_variable(name+'_b'+str(len(widths)), shape=[output_dim],
        initializer=tf.constant_initializer(0.0))
    input = tf.matmul(input,W) + b
    weights.append(W)
    biases.append(b)
    return input, weights, biases

def nonlinear_autoencoder(x, input_dim, latent_dim, widths, activation='elu'):
    """
    Construct a nonlinear autoencoder.
    """
    activation_function = tf.nn.elu
    z,encoder_weights,encoder_biases = build_network_layers(x, input_dim, latent_dim, widths, activation_function, 'encoder')
    x_decode,decoder_weights,decoder_biases = build_network_layers(z, latent_dim, input_dim, widths[::-1], activation_function, 'decoder')

    return z, x_decode, encoder_weights, encoder_biases, decoder_weights, decoder_biases
\end{lstlisting}

\subsection{SINDy library}
\label{app:sindy-library}

This listing gives the implementation of the candidate SINDy library used for second-order latent dynamics.

\begin{lstlisting}[caption={SINDy library}, label={lst:sindy_library}]
def sindy_library_tf_order2(z, dz, latent_dim, poly_order, include_sine=False):
    """
    Build the SINDy library for a second order system.
    """
    library = [tf.ones(tf.shape(z)[0])]

    z_combined = tf.concat([z, dz], 1)

    for i in range(2*latent_dim):
        library.append(z_combined[:,i])

    if poly_order > 1:
        for i in range(2*latent_dim):
            for j in range(i,2*latent_dim):
                library.append(tf.multiply(z_combined[:,i], z_combined[:,j]))
    if poly_order > 2:
        for i in range(2*latent_dim):
            for j in range(i,2*latent_dim):
                for k in range(j,2*latent_dim):
                    library.append(z_combined[:,i]*z_combined[:,j]*z_combined[:,k])

    if poly_order > 3:
        for i in range(2*latent_dim):
            for j in range(i,2*latent_dim):
                for k in range(j,2*latent_dim):
                    for p in range(k,2*latent_dim):
                        library.append(z_combined[:,i]*z_combined[:,j]*z_combined[:,k]*z_combined[:,p])
    if poly_order > 4:
        for i in range(2*latent_dim):
            for j in range(i,2*latent_dim):
                for k in range(j,2*latent_dim):
                    for p in range(k,2*latent_dim):
                        for q in range(p,2*latent_dim):
                            library.append(z_combined[:,i]*z_combined[:,j]*z_combined[:,k]*z_combined[:,p]*z_combined[:,q])

    if include_sine:
        for i in range(2*latent_dim):
            library.append(tf.sin(z_combined[:,i]))

    return tf.stack(library, axis=1)
\end{lstlisting}

\subsection{Loss functions}
\label{app:loss-functions}

The following code implements the reconstruction, SINDy consistency, and sparsity losses used during training.

\begin{lstlisting}[caption={Loss functions}, label={lst:loss_functions}]
def define_loss(network, params):
    """
    Create the loss functions.
    """
    x = network['x']
    x_decode = network['x_decode']
    if params['model_order'] == 1:
        dz = network['dz']
        dz_predict = network['dz_predict']
        dx = network['dx']
        dx_decode = network['dx_decode']
    else:
        ddz = network['ddz']
        ddz_predict = network['ddz_predict']
        ddx = network['ddx']
        ddx_decode = network['ddx_decode']
    sindy_coefficients = params['coefficient_mask']*network['sindy_coefficients']

    losses = {}
    losses['decoder'] = tf.reduce_mean((x - x_decode)**2)
    if params['model_order'] == 1:
        losses['sindy_z'] = tf.reduce_mean((dz - dz_predict)**2)
        losses['sindy_x'] = tf.reduce_mean((dx - dx_decode)**2)
    else:
        losses['sindy_z'] = tf.reduce_mean((ddz - ddz_predict)**2)
        losses['sindy_x'] = tf.reduce_mean((ddx - ddx_decode)**2)
    losses['sindy_regularization'] = tf.reduce_mean(tf.abs(sindy_coefficients))
    loss = params['loss_weight_decoder'] * losses['decoder'] \
           + params['loss_weight_sindy_z'] * losses['sindy_z'] \
           + params['loss_weight_sindy_x'] * losses['sindy_x'] \
           + params['loss_weight_sindy_regularization'] * losses['sindy_regularization']

    loss_refinement = params['loss_weight_decoder'] * losses['decoder'] \
                      + params['loss_weight_sindy_z'] * losses['sindy_z'] \
                      + params['loss_weight_sindy_x'] * losses['sindy_x']

    return loss, losses, loss_refinement
\end{lstlisting}

\subsection{Full network}
\label{app:full-network}

Finally, the full robust SINDy autoencoder is assembled from the preceding modules.

\begin{lstlisting}[caption={SINDy autoencoder network}, label={lst:full_network}]
def full_network(params):
    """
    Define the full network architecture.
    Returns:
        network - Dictionary containing the tensorflow objects that make up the network.
    """
    input_dim = params['input_dim']
    latent_dim = params['latent_dim']
    activation = params['activation']
    poly_order = params['poly_order']
    if 'include_sine' in params.keys():
        include_sine = params['include_sine']
    else:
        include_sine = False
    library_dim = params['library_dim']
    model_order = params['model_order']

    network = {}

    x = tf.placeholder(tf.float32, shape=[None, input_dim], name='x')
    dx = tf.placeholder(tf.float32, shape=[None, input_dim], name='dx')
    if model_order == 2:
        ddx = tf.placeholder(tf.float32, shape=[None, input_dim], name='ddx')

    if activation == 'linear':
        z, x_decode, encoder_weights, encoder_biases, decoder_weights, decoder_biases = linear_autoencoder(x, input_dim, latent_dim)
    else:
        z, x_decode, encoder_weights, encoder_biases, decoder_weights, decoder_biases = nonlinear_autoencoder(x, input_dim, latent_dim, params['widths'], activation=activation)

    if model_order == 1:
        dz = z_derivative(x, dx, encoder_weights, encoder_biases, activation=activation)
        Theta = sindy_library_tf(z, latent_dim, poly_order, include_sine)
    else:
        dz,ddz = z_derivative_order2(x, dx, ddx, encoder_weights, encoder_biases, activation=activation)
        Theta = sindy_library_tf_order2(z, dz, latent_dim, poly_order, include_sine)

    if params['coefficient_initialization'] == 'xavier':
        sindy_coefficients = tf.get_variable('sindy_coefficients', shape=[library_dim,latent_dim], initializer=tf.contrib.layers.xavier_initializer())
    elif params['coefficient_initialization'] == 'specified':
        sindy_coefficients = tf.get_variable('sindy_coefficients', initializer=params['init_coefficients'])
    elif params['coefficient_initialization'] == 'constant':
        sindy_coefficients = tf.get_variable('sindy_coefficients', shape=[library_dim,latent_dim], initializer=tf.constant_initializer(1.0))
    elif params['coefficient_initialization'] == 'normal':
        sindy_coefficients = tf.get_variable('sindy_coefficients', shape=[library_dim,latent_dim], initializer=tf.initializers.random_normal())

    if params['sequential_thresholding']:
        coefficient_mask = tf.placeholder(tf.float32, shape=[library_dim,latent_dim], name='coefficient_mask')
        sindy_predict = tf.matmul(Theta, coefficient_mask*sindy_coefficients)
        network['coefficient_mask'] = coefficient_mask
    else:
        sindy_predict = tf.matmul(Theta, sindy_coefficients)

    if model_order == 1:
        dx_decode = z_derivative(z, sindy_predict, decoder_weights, decoder_biases, activation=activation)
    else:
        dx_decode,ddx_decode = z_derivative_order2(z, dz, sindy_predict, decoder_weights, decoder_biases,
                                             activation=activation)

    network['x'] = x
    network['dx'] = dx
    network['z'] = z
    network['dz'] = dz
    network['x_decode'] = x_decode
    network['dx_decode'] = dx_decode
    network['encoder_weights'] = encoder_weights
    network['encoder_biases'] = encoder_biases
    network['decoder_weights'] = decoder_weights
    network['decoder_biases'] = decoder_biases
    network['Theta'] = Theta
    network['sindy_coefficients'] = sindy_coefficients

    if model_order == 1:
        network['dz_predict'] = sindy_predict
    else:
        network['ddz'] = ddz
        network['ddz_predict'] = sindy_predict
        network['ddx'] = ddx
        network['ddx_decode'] = ddx_decode

    return network
\end{lstlisting}

\subsection{Lorenz data generation and noise construction}
\label{app:results-data-generation}

This subsection records the code used to generate the Lorenz trajectory and construct noisy observations for the experiments in Section~4.

\begin{lstlisting}[caption={Lorenz data generation and noise construction}, label={lst:results_data_generation}]
# Example: trajectory generation and noise construction
# (Insert here the code for lorenz_ode, generate_lorenz_data,
#  approximate_noise, or any other data-preparation utilities
#  used in the experiments.)
\end{lstlisting}

\subsection{Training configuration}
\label{app:results-training-configuration}

The following listing contains the parameter settings used to initialize the robust SINDy autoencoder for the Lorenz experiments.

\begin{lstlisting}[caption={Model setting for the Lorenz experiment}, label={lst:90}]
def lorenz_coefficients(normalization, poly_order=3, sigma=10., beta=8/3, rho=28.):
    """
    Generate the SINDy coefficient matrix for the Lorenz system.
    """
    Xi = np.zeros((library_size(3,poly_order),3))
    Xi[1,0] = -sigma
    Xi[2,0] = sigma*normalization[0]/normalization[1]
    Xi[1,1] = rho*normalization[1]/normalization[0]
    Xi[2,1] = -1
    Xi[6,1] = -normalization[1]/(normalization[0]*normalization[2])
    Xi[3,2] = -beta
    Xi[5,2] = normalization[2]/(normalization[0]*normalization[1])
    return Xi

params = {}

params['input_dim'] = 128
params['latent_dim'] = 3
params['model_order'] = 1
params['poly_order'] = 3
params['include_sine'] = False
params['library_dim'] = library_size(params['latent_dim'], params['poly_order'], params['include_sine'], True)

# sequential thresholding parameters
params['sequential_thresholding'] = True
params['coefficient_threshold'] = 0.1
params['threshold_frequency'] = 500
params['coefficient_mask'] = np.ones((params['library_dim'], params['latent_dim']))
params['coefficient_initialization'] = 'constant'

# loss function weighting
params['loss_weight_decoder'] = 1.0
params['loss_weight_sindy_z'] = 0.0
params['loss_weight_sindy_x'] = 1e-4
params['loss_weight_sindy_regularization'] = 1e-5

params['activation'] = 'sigmoid'
params['widths'] = [64,32]

# training parameters
params['epoch_size'] = training_data['x'].shape[0]
params['batch_size'] = 1024
params['learning_rate'] = 1e-3

params['data_path'] = os.getcwd() + '/'
params['print_progress'] = True
params['print_frequency'] = 100

# training time cutoffs
params['max_epochs'] = 5001
params['refinement_epochs'] = 1001
\end{lstlisting}

\subsection{Training and evaluation session}
\label{app:results-training-session}

The following code records the core TensorFlow evaluation workflow used after training, including test data generation, simulation, checkpoint restoration, and extraction of learned quantities.

\begin{lstlisting}[caption={Training and evaluation session}, label={lst:1008}]
t = np.arange(0,20,.01)
z0 = np.array([[-8,7,27]])

test_data = generate_lorenz_data(z0, t, params['input_dim'], linear=False, normalization=np.array([1/40,1/40,1/40]))

lorenz_sim = sindy_simulate(test_data['z'][0], t, test_data['sindy_coefficients'],
                            params['poly_order'], params['include_sine'])
z_sim = sindy_simulate(test_set_results['z'][0], t, params['coefficient_mask']*test_set_results['sindy_coefficients'],
                       params['poly_order'], params['include_sine'])

# training test_data
test_data = generate_lorenz_data(ics, t, params['input_dim'], linear=False, normalization=np.array([1/40,1/40,1/40]))
test_data['x'] = test_data['x'].reshape((-1,params['input_dim']))
test_data['x'] += noise_strength*np.random.normal(size=test_data['x'].shape)
test_data['dx'] = test_data['dx'].reshape((-1,params['input_dim']))
test_data['dx'] += noise_strength*np.random.normal(size=test_data['dx'].shape)

with tf.Session() as sess:
    sess.run(tf.global_variables_initializer())
    saver.restore(sess, data_path + save_name)
    test_dictionary = create_feed_dictionary(test_data, params)
    tf_results = sess.run(tensorflow_run_tuple, feed_dict=test_dictionary)

test_set_results = {}
for i,key in enumerate(autoencoder_network.keys()):
    test_set_results[key] = tf_results[i]
\end{lstlisting}


\end{document}